\documentclass[epj]{svjour}
\usepackage{graphicx}
\begin{document}
\title{Electron-doping versus hole-doping in the 2D $t$-$t'$
Hubbard model}

\author{Carsten Honerkamp}

\institute{ Theoretische Physik,
ETH-H\"onggerberg, CH-8093 Z\"urich, Switzerland}
\date{\today}

\abstract{We compare the one-loop renormalization group flow to strong coupling of the electronic interactions in the
two-dimensional  $t$-$t'$-Hubbard model with $t'=-0.3t$ for band fillings smaller and larger than half-filling.
Using a numerical $N$-patch scheme ($N=32\dots 96$) we show that in the
electron-doped case with decreasing electron density there is a rapid transition 
from a $d_{x^2-y^2}$-wave superconducting
regime with small characteristic energy scale to an approximate nesting regime
with strong antiferromagnetic tendencies and higher energy scales.  This 
contrasts with the hole-doped side discussed recently which
exhibits a broad parameter region where the renormalization group flow suggests a truncation of
the Fermi surface at the saddle points. We compare the quasiparticle scattering rates
obtained from the renormalization group calculation which further emphasize the
differences between the two cases. 
} 
\authorrunning{C. Honerkamp} 
\titlerunning{Electron- versus hole-doping in the $t$-$t'$
Hubbard model}
\maketitle
 
\section{Introduction}
Viewed from a weak coupling perspective, the two-dimen\-sional Hubbard model exhibits
an interesting interplay between different types of fluctuations. Particular
attention has been devoted to the situation close to half band filling, mainly because of its
relevance to the high-$T_c$ superconducting cuprates \cite{anderson}. 
For these particle densities, because of the
approximate nesting between opposite sides of the Fermi surface (FS) or due to
the large density of states
around the saddle points of the dispersion at $(\pi,0) $ and $(0, \pi)$, the
scattering processes with momentum transfer $\approx (\pi,\pi)$ will be
strongly enhanced. Besides the fact that these scattering processes drive antiferromagnetic
(AF) fluctuations, they also create a $d_{x^2-y^2}$-wave component 
in the pair scattering \cite{scalapino}. 
One-loop renormalization group (RG) techniques represent 
a powerful method to analyze the coupling between antiferromagnetic and superconducting
fluctuations in an 
unbiased way (see
\cite{dzialoshinski,schulz,lederer,zanchi,furukawa,halboth,hmprl,honerkamp,tsai} and 
the present paper).
 
Apart from the 
competition between two different types of ground states, the hole-doped $t$-$t'$ Hubbard
model close to half filling may exhibit a richer variety of possible
ground states especially when umklapp scattering between electrons becomes
important. In \cite{furukawa} and \cite{honerkamp} it was argued that in the
case of not-too-small values of the next-nearest neighbor hopping $t'$, e.g. $t'=-0.3t$, when the FS 
is close to the saddle points at $(\pi,0)$ and $(0,\pi)$, the RG
flow to strong coupling between the $\vec{k}$-space regions around the saddle
points has strong similarities with the RG flow in the
half-filled two-leg Hubbard ladder. In the latter system, the ground state is well
understood and is an insulating spin liquid (ISL) \cite{fisher}. Although there is no
reliable theory for the two-dimensional case, the similarity of the RG
flows suggests that here as well the true strong coupling 
state of the $\vec{k}$-space regions around the saddle points is an ISL with a truncated FS. The parameter
region in which this interesting flow to strong coupling occurs was called
the {\em saddle point regime}. 

In this paper we investigate the RG flow to strong coupling for
band fillings larger than half-filling, corresponding to electron-doping the
half-filled system. Again we choose $t'=-0.3t$. The main observation is that in the
electron-doped case the saddle point
regime of the hole-doped case is absent. Instead of the mutually reinforcing
flow between AF and $d$-wave pairing processes which was found in the saddle
point regime, there is a 
clear separation between the energy scales of the two channels. 
We do not find any signs
for a Fermi surface truncation around the saddle points. Moreover the ''hot''
Brillouin zone (BZ)
regions responsible for the leading flow are now located closer to the BZ
diagonals. The different result 
is a direct consequence of the location of the Fermi surface,
which in the electron-doped case crosses the umklapp surface (US, defined in
Fig. \ref{geom}) in the BZ diagonal, therefore giving rise to  
strong scattering between the FS segments on opposite sides connected by the
wave vector $(\pi, \pi)$. These AF processes generate only 
a weak attractive $d_{x^2-y^2}$ component in the
pair scattering channel and do not drive the pairing processes at low scales
as they do in the saddle point regime of the hole-doped case. Consequently, upon
increasing the band filling such that the $(\pi, \pi)$ processes get cut off, 
we obtain a rather abrupt transition from a flow
to strong 
coupling dominated by umklapp and AF processes with high critical scale to a 
Kohn-Luttinger-like instability with predominant $d_{x^2-y^2}$-Cooper
processes at comparably low critical scale. Therefore the electron-doped case considered 
here 
strongly resembles the hole-doped cases with small absolute values of $t'$
analyzed in \cite{zanchi} and \cite{halboth,hmprl} and differs from the hole-doped case
with $t'=-0.3t$ discussed in \cite{honerkamp}.

As another piece of information we present results for the scattering rates of
quasiparticles at the FS with a  RG method described below. These calculations 
further illustrate
the differences between the hole- and electron-doped cases.
According to model calculations by Ioffe and Millis \cite{ioffe} and also Hlubina \cite{hlubina} most transport experiments of the optimally hole-doped cuprates can
be properly described by assuming a pronounced anisotropy in the scattering
rate. The anisotropy is such that the quasiparticles around the saddle
points scatter strongly and their spectral weight becomes smeared out while
the quasiparticles in the BZ diagonal are only subject to a weak Fermi-liquid-like
 scattering. Our RG calculations with frequency-independent coupling constants can certainly not give reliable
information about the frequency dependence of the selfenergy. On the other hand their $\vec{k}$-space resolution is rather good. Applying the RG
scheme we observe a pronounced angular dependence of $\mathrm{Im} \Sigma (
\vec{k}_F,\omega=0)$ with maximal scattering for particles at the saddle
points and weaker scattering in the BZ diagonals.   In the electron-doped
case, the anisotropy is weaker and the FS regions with the largest scattering
rate are located in the BZ diagonals, tied to the intersection of the FS with
the US.

Angular resolved photoemission (ARPES) allows one to measure the temperature
dependence of Im $\Sigma (\omega=0,\vec{k}_F)$ directly from the width of the
quasiparticle peaks. For optimally hole-doped Bi 2212, Valla et al. \cite{valla}
found a linear-$T$ dependence almost everywhere on  the FS. For a comparison
we calculate the quasiparticle scattering rates at high temperatures, where
the RG flow does not diverge. Our results for the hole-doped case are qualitatively similar to the
experimental data. In the electron-doped case the RG calculations
yield a non-linear $T$-dependence of the scattering rates closer to
conventional $T^2$ behavior.

\section{The method}
Here we describe the RG method and the scheme for the calculation of the
quasiparticle scattering rates. Readers who are only interested in the results 
can proceed to the next section.
\subsection{RG flow of interactions and susceptibilities}
The interplay between the different fluctuations mentioned 
 in the introduction can be appropriately analyzed within
one-loop RG where one successively integrates out intermediate
states according to their band energy $\epsilon_\vec{k}$. For the two-dimen\-sional Hubbard
model on the square lattice with nearest neighbor hopping $t$ and
next-nearest neighbor hopping $t'$,
\[ \epsilon_\vec{k} = -2 t \left( \cos k_x + \cos k_y \right) - 4t'  \cos k_x
 \cos k_y - \mu \, , \]
where $\mu$ is the chemical potential and the lattice constant is set to
unity. 

The RG scheme for 1PI irreducible vertex
functions which we use is explained and discussed in
detail in Ref. \cite{salmhofer}. Here we give simple graphical
explanation of the RG flow for the two- and four-point vertex
functions. The flow of higher order irreducible vertex functions is not taken into account.

\begin{figure}
\begin{center} 
\includegraphics[width=.45\textwidth]{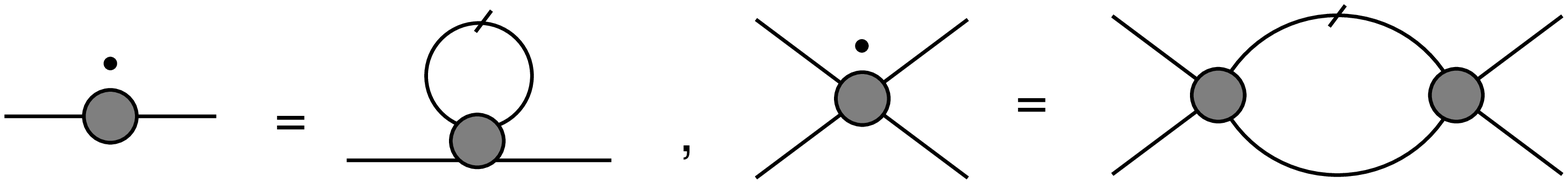}
\end{center} 
\caption{The one-loop RG equations for two - and four-point vertex
function. The dot symbolizes the derivative with respect to the energy scale $\Lambda$.}
\label{rgdia24}
\end{figure}

The differential equations for the two-point vertex (yielding the scale-dependent selfenergy
$\Sigma_\Lambda (\vec{k},i\omega)$) and the four-point vertex function, which
describe their RG flow with decreasing energy scale $\Lambda$, are
described graphically in Fig. \ref{rgdia24}. At the internal vertices we have four-point vertex functions at
scale $\Lambda$ connecting the incoming and outgoing lines with the
internal ones. Crystal momenta and Matsubara frequencies are conserved at the vertices. One of the two
internal lines corresponds to a so-called single scale propagator $S_\Lambda$ which is only non-zero for modes at
scale $\Lambda$ while the other line denotes a full Green's function
$G_\Lambda$ with low
energy modes below scale $\Lambda$ cut out through a cutoff function
$C(\Lambda, |\epsilon_\vec{k}|)$.
 Throughout this paper we neglect
possible self-energy corrections, therefore we have
\[ G_\Lambda (\vec{k}, i \omega) = \frac{C(\Lambda, |\epsilon_\vec{k}|)}{ i
\omega - \epsilon_\vec{k} } \quad \mbox{and}  \quad S_\Lambda (\vec{k}, i \omega) =
\frac{\partial_\Lambda C(\Lambda, |\epsilon_\vec{k}|)}{ i
\omega - \epsilon_\vec{k} } \, . \]

\begin{figure} 
\begin{center}
\includegraphics[width=.15\textwidth]{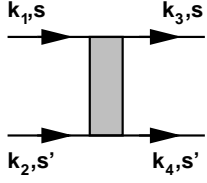} \end{center}
\caption{The coupling function $V_\Lambda
(\vec{k}_1,\vec{k}_2,\vec{k}_3)$. $\vec{k}_1,s$ ($\vec{k}_2,s'$) specify
the first (second) incoming particle, $\vec{k}_3,s$ ($\vec{k}_4,s'$) belong 
to the first (second) outgoing particle.}
\label{vlambda}
\end{figure}

\begin{figure} 
\begin{center}
\includegraphics[width=.34\textwidth]{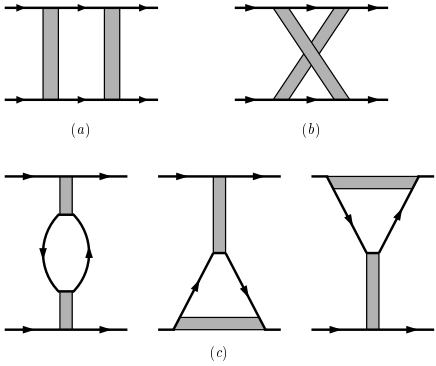} \end{center}
\caption{The contributions to the right--hand side of the RG equation for the
coupling function $V_\Lambda
(\vec{k}_1,\vec{k}_2,\vec{k}_3)$. 
$(a)$ the particle--particle term
$(b)$ the crossed particle--hole term
$(c)$ the direct particle--hole terms; the first of these three graphs gets 
a factor $-2$ because of the fermion loop.}
\label{diagrams}
\end{figure}

The scale-dependent spin-rotation invariant four-point vertex function
$\Gamma_\Lambda ( \vec{k}_1,s_1,\vec{k}_2,s_2,\vec{k}_3,s_3,\vec{k}_4,s_4)$
can be expressed in terms of a coupling function
$V_\Lambda (\vec{k}_1,\vec{k}_2,\vec{k}_3)$ \cite{honerkamp,salmhofer}
which is determined for the case $s_1=s_3$ and
$s_2=s_4$. The $s_i$ denote the $z$-components of the spins of incoming ($s_1$ 
and $s_2$) and
outgoing ($s_3$ and $s_4$) particles. $V_\Lambda
(\vec{k}_1,\vec{k}_2,\vec{k}_3)$ is represented
graphically in Fig. \ref{vlambda}. This means that the electron-electron
interaction at energy scale $\Lambda$  is 
a function of the two incoming wave vectors $\vec{k}_1$ and $\vec{k}_2 $ and
one outgoing wave vector $\vec{k}_3$.   For the bare Hubbard interaction at starting scale
$\Lambda_0 \approx 4t$, we set $V_{\Lambda_0}
(\vec{k}_1,\vec{k}_2,\vec{k}_3)=U$. For all calculations in this paper we
choose $U=3t$. The flow of $V_\Lambda
(\vec{k}_1,\vec{k}_2,\vec{k}_3)$ with decreasing the energy scale $\Lambda$ is 
then
given by all types of one-loop diagrams including particle-particle and 
particle-hole diagrams with all possible momentum transfers (see
Fig. \ref{diagrams}).  

For the numerical integration of these coupled equations we use
a phase space discretization following Zanchi and Schulz \cite{zanchi}.  The
BZ is divided up into $nN$ patches centered around $N$ lines.  
Each line starts from the 
origin in a certain angular direction $\theta(k)$ and 
from the $(\pm \pi ,\pm \pi)$-points so that the lines meet at the umklapp 
surface.  The phase space segments around  these lines are then further 
split into $n$ patches. Here we took $n=3$ where one patch per line is
centered around the FS with e.g. $|\epsilon(\vec{k})|< 0.4t$, while the other two patches cover phase space regions
at higher positive or negative band energy. 

\begin{figure} 
\begin{center} 
\includegraphics[width=.32\textwidth]{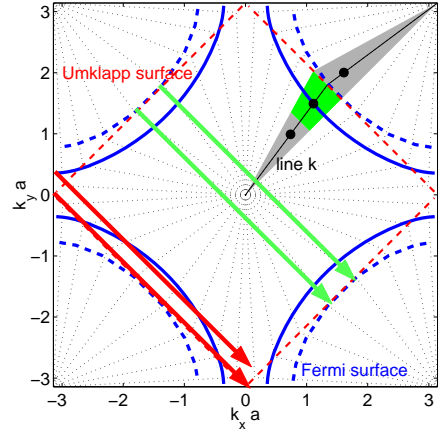}
\end{center}
\caption{The Brillouin zone, Fermi and umklapp surfaces and the lines in the
patch centers for $N= 32$. The solid dots in the patches denote the wave vectors for 
which the four-point vertex function is calculated. For all wave vectors
inside the same 
patch, $V_\Lambda (\vec{k}_1,\vec{k}_2,\vec{k}_3)$ is approximated by a
constant. The solid line denotes the non-interacting FS for $t'=-0.3t$ and $\mu=-t$, which is
in the saddle point regime of hole-doped case. The curved dashed line is a
typical FS for the
electron-doped case. The straight dashed lines connecting the $(\pm \pi, 0)$- and
$(0,\pm \pi)$-points denote the umklapp surface. The arrows symbolize umklapp
processes between the saddle point and the BZ diagonals, respectively. }
\label{geom}
\end{figure}

\begin{figure} 
\begin{center} 
\includegraphics[width=.5\textwidth]{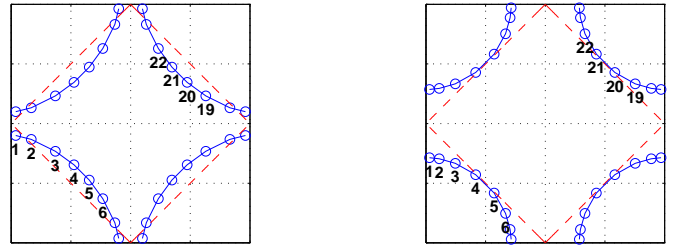}
\end{center}
\caption{The Fermi surfaces and $N=32$ points $\vec{k}_F (k_i)$ for which the coupling function
$V_\Lambda (k_1,k_2,k_3)$ is calculated. Left plot: $\mu=-1.05t$
(hole-doped, $\langle n \rangle
\approx 0.83$). Right plot: $\mu = 0$ (electron-doped, $\langle n \rangle
\approx 1.2$ per site).}
\label{fshded}
\end{figure}

Then the coupling function is discretized as follows: we approximate $V_{\Lambda} (\vec{k}_1,\vec{k}_2,\vec{k}_3)$ by a constant for all wave
vectors in the same patches and calculate the RG flow for the
subset of interaction vertices with one wavevector representative for each
patch. In the low energy patches around the FS we choose to take these wave vectors as the
crossing points of the $N$ lines with the Fermi surface (FS), the wave vectors for 
the remaining patches away from the FS correspond to a higher band energy,
e.g. $|\epsilon(\vec{k})|=0.8t$ (see Fig. \ref{geom}). The phase space space integrations
are performed as sums over the patches and integrations over the radial 
direction along 3 or 5 lines inside each patch. Most calculations were done
using a 32$\times$3 system, i.e. $N=32$ and $n=3$. 
Two typical Fermi surfaces with $N=32$ points 
are shown in Fig. \ref{fshded}. 

Along with the flow of the interactions we calculate several static
susceptibilities by coupling external fields of appropriate form to the
electrons. During the flow these external coupling are renormalized through
one-loop vertex corrections, as described in Ref. \cite{honerkamp}. The pairing and AF susceptibilities can be calculated directly
with the RG flow of the coupling function. The
susceptibilities and couplings to uniform external charge and spin fields
require special care as they only involve particle-hole excitations in a shell 
around the FS 
with width of the temperature. Before the cutoff $\Lambda$ reaches that
scale the interactions have typically grown larger than the bandwidth such
that the one-loop flow is not reliable any longer. In order to get a
qualitative result for these cases we stop the flow of the interactions at a
certain scale $\Lambda_{\mathrm{freeze}}$ and treat them as constants in the continuation of the flow of
the susceptibilities down to $\Lambda =0$ where we obtain the main contributions to the uniform
susceptibilities.
This is basically equivalent to stopping the RG flow at $\Lambda_{\mathrm{freeze}}$ and
calculating these susceptibilities for the effective
theory below the cutoff $\Lambda_{\mathrm{freeze}}$  within RPA (random phase
approximation) using the renormalized
interactions given by the RG down to $\Lambda_{\mathrm{freeze}}$. A similar way
to obtain the uniform susceptibilities from the flow of the Landau function
has been used in Ref. \cite{halboth}.

\subsection{Approximate calculation of the scattering rates}
\label{scatratform}
Here we describe an approximation scheme for the calculation of the imaginary
part of the selfenergy from the flow of coupling function. 
The technical problem occurring in the RG formalism presented above is that we have neglected all frequency dependencies of the four-point
vertex, hence the  four-point vertex in the one-loop contribution to $\Sigma$ at scale $\Lambda$ in Fig. \ref{rgdia24} 
 is a real quantity and the diagram does not give an
imaginary part. This deficiency can be repaired to a large extent by replacing
the approximate frequency-independent vertex with the solution of the one-loop
RG equation for the four-point vertex in its integral form. This effectively yields a two-loop term corresponding to a
two-particle--one-hole intermediate state which, after analytical
continuation to real frequencies, gives a
non-vanishing imaginary part. Moreover these contributions to
Im $\Sigma_\Lambda$ are the
leading ones, because all neglected contributions from further insertions of one-loop diagrams at the vertices
have intermediate states with a number of intermediate particles larger than
3 and can therefore be expected to give only small contributions due to phase space
restrictions. If we ignored the flow of the coupling
constants and integrated the expression thus obtained down to $\Lambda = 0$, we
would retrieve the bare two-loop selfenergy. 
In this paper we neglect the
feedback of the selfenergy on the flow of the coupling function. In
principal this effect is contained in the RG equations but keeping it would
increase the complexity of the calculations considerably.

Formally we can calculate the selfenergy $\Sigma_{\Lambda =0}  (\vec{k}_F ,
i\omega) $ at the FS for Matsubara frequency $i \omega$  by integrating the RG differential
equation for the scale-dependent selfenergy depicted in Fig.
\ref{rgdia24} as
follows:
\begin{eqnarray*} \Sigma_{\Lambda =0} (\vec{k}_F , i\omega) = \int_{\Lambda_0}^0 d\Lambda \,
\int \frac{d^2k'}{(2\pi)^2} \sum_{i\omega'} \, \qquad \ \quad \qquad  \\  \qquad
\left[ -2 V_\Lambda(\vec{k}_F,\vec{k}',\vec{k}_F) +
V_\Lambda(\vec{k}_F,\vec{k}',\vec{k}') \right] S_\Lambda (\vec{k}', i
\omega') \end{eqnarray*}
For $V_\Lambda$ we now substitute the differential equation for the
four-point vertex, integrated from the starting scale $\Lambda_0$ down to
$\Lambda$.
This corresponds to inserting the one-loop diagram on the right side of
Fig. \ref{rgdia24} into the self-energy graph on the left side. Then we get a
second integral over $d\Lambda'$:
\begin{eqnarray}  \Sigma_{\Lambda =0} (\vec{k}_F , i\omega) =\hspace{-1mm} 
\int_{\Lambda_0}^0 \hspace{-1.5mm} d\Lambda \hspace{-1mm} 
\, \int_{\Lambda_0}^{\Lambda} \hspace{-1.5mm} d\Lambda' \, \sum V_{\Lambda'} G_{\Lambda'} S_{\Lambda'} S_{\Lambda}
V_{\Lambda'} \label{doubleint} \end{eqnarray}
Here the sum is over all internal wavevectors, frequencies and spin indices of 
the different  possible diagrams. 
For simplicity we have suppressed the arguments
of the four-point vertices and propagators in this expression. 
Obviously, repeating the insertion of
scale-integrated one-loop terms for the vertices yields a perturbation
expansion of the self-energy with arbitrary numbers of intermediate states.
Here we restrict the analysis to the contributions from the two-loop term
(\ref{doubleint}), which contains the Landau-Fermi liquid selfenergy and the
deviations from it through the flow to strong coupling of the coupling
function $V_\Lambda$.

In the integrand in (\ref{doubleint}), the two four-point
vertices $V_{\Lambda'}$, one single scale propagator $S_{\Lambda'}$ and one full propagator
$G_{\Lambda'} $ depend on $\Lambda'$. The only $\Lambda$-dependence
is in the differentiated cutoff-function $\dot{C}(\Lambda)$ of the original single-scale propagator $S_{\Lambda} $.  Therefore (\ref{doubleint}) has
the structure 
\[  \Sigma_{\Lambda =0} (\vec{k}_F , i\omega) = \sum\int_{\Lambda_0}^0 d\Lambda
\, \dot{C}(\Lambda) \, \int_{\Lambda_0}^\Lambda
d\Lambda'  \, \, R(\Lambda') \, 
. \] 
 The double scale integral is numerically expensive, but we can circumvent it by a
partial integration, resulting in
\begin{eqnarray} \Sigma_{\Lambda =0} (\vec{k}_F , i\omega) =  \sum\int_{\Lambda_0}^0 d\Lambda  \,  \left[ C(0) -
C(\Lambda) \right] 
\,  R(\Lambda)  \,  . \label{respartint}\end{eqnarray}
The remainder $ R(\Lambda)$ contains one single scale propagator at $\Lambda$
and a propagator for modes above $\Lambda$. The difference of cutoff
functions in the rectangular brackets means that the third of the three
internal lines has its support on modes below the cutoff. The form
(\ref{respartint}) only contains a single $\Lambda$-integral and the vertex
functions at $\Lambda$. Therefore it can be integrated along with the flow of
the four-point vertex. Note however that for consistency with
(\ref{doubleint}) we should always integrate out the full scale range.

\begin{figure}
\begin{center} 
\includegraphics[width=.45\textwidth]{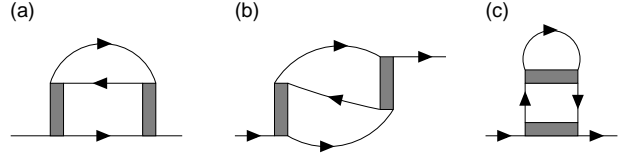}
\end{center} 
\caption{The different diagrams for the two-loop selfenergy. Diagrams (a) and (b) contribute to the imaginary part of the selfenergy, contributions of
type (c) are real for external frequency $\omega +i \delta$. There are 3 other diagrams of type (c) with different
orientation of the vertices which are not shown here. }
\label{3_2loops}
\end{figure}

Diagrammatically the insertion of the one-loop flow into the one-loop diagram
for the selfenergy yields three topologically different diagrams which are
shown in Fig. \ref{3_2loops}. Each of these three diagrams gives 6
contributions: one of the internal line contains modes above
$\Lambda$, another modes at $\Lambda$, and the third below $\Lambda$. 

Since we are interested in the quasiparticle scattering rates, we perform an 
analytical continuation onto the real frequency axis: the integrand of the two 
diagrams (a) and (b) in Fig. \ref{3_2loops} contains the factor
\[ \frac{1}{i \omega - \epsilon_1 -\epsilon_2 + \epsilon_3} \]
which upon replacing $i \omega \to \omega + i \delta$ with $\omega=0$ yields
an imaginary part $\propto \delta ( \epsilon_1 + \epsilon_2 -
\epsilon_3)$. The $\delta$-function is smeared with a small width for the numerical treatment. 

One can show that for scale-independent vertices, e.g. all $V_\Lambda (\vec{k}_1,\vec{k}_2,\vec{k}_3) =U$, this scheme is equivalent to the
calculation of the bare two-loop diagram. Thus for ordinary cases where the
flow does not diverge we should obtain two-dimensional Fermi-liquid results. In our case the flow goes to
strong coupling, and this leaves two possibilities for the analysis: either we
choose a high temperature such that the couplings do not become too large, then
we can apply the above scheme down to zero scale and obtain an estimate for
the imaginary part of the selfenergy above the strong coupling phase.  The
other option is, if we want to analyze the situation in the strong coupling
regime, to stop the flow of the couplings at some scale
$\Lambda_{\mathrm{freeze}}$ and integrate the
flow down to zero scale with fixed couplings. We can then vary
$\Lambda_{\mathrm{freeze}}$ and thus obtain the change in the imaginary part of 
the selfenergy due to the flow of the couplings. This will tell us on which FS 
parts the renormalized interactions weaken the quasiparticles most and we will e.g. find
that in the hole-doped case the quasiparticles in the BZ diagonals are not affected by strong
scattering. The results of both types of analysis are described in Section \ref{scatrat}.

\section{Comparison of the flows to strong coupling}
First we compare the RG flow of the interactions. For the chosen
parameters we generally find a flow to strong coupling, i.e. at sufficiently
low scales and temperatures some components
of the coupling function $V_\Lambda (k_1,k_2,k_3)$ take absolute values larger than the
band energy. We analyze the flow to
strong coupling at scale $\Lambda_W$ where the coupling functions just exceed
the order of the bare bandwidth, i.e. at $V_{\Lambda_W, \mathrm{max.}} \approx 8-12t$. At these scales and for typical temperatures, the coupling functions have
already developed a pronounced $\vec{k}$-space structure and the dominant 
interaction terms at this scale will certainly be important for the
strong coupling state. On the other hand
the FS shift due to
one-loop selfenergy corrections is still small \cite{honerkamp} such that it
does not qualitatively change the flow above $\Lambda_W$. Similarly 
the scattering rate
for particles at the FS (see below) remains smaller than $\Lambda_W$. Note
however that at $\Lambda_W$ and for initial interaction $U=3t$ the flow has not reached an asymptotic
form and different classes of coupling constants would evolve differently if we
continued the flow below $\Lambda_W$ (where our method breaks down). Therefore the analysis of the flow to strong coupling remains
qualitative and does not provide definitive conclusions about the true strong
coupling state. 

\subsection{Hole-doped model}
\begin{center} 
\begin{figure} 
\includegraphics[width=.48\textwidth]{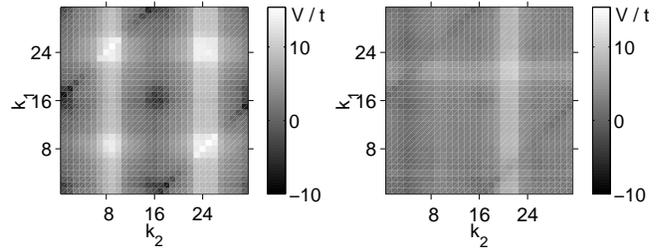}
\caption{Snapshot of the RG flow  when the couplings have reached values
larger than the bandwidth for the hole-doped case in the saddle point regime, $t'=-0.3t$ and
$\mu=-t$. The colorbar denotes the value of the couplings $V(k_1,k_2,k_3)/t$,
where $\vec{k}_F(k_1)$ and $\vec{k}_F(k_2)$ are the two incoming wavevectors
on the FS and the outgoing wavevector $\vec{k}_F(k_3)$ is
fixed at point 1 or 4. The 32 points are numbered according to their position
around the FS, points 1,8,9,16 etc. are closest to the saddle points.}
\label{hdvcomp}
\end{figure}
\end{center}

\begin{center} 
\begin{figure} 
\includegraphics[width=.48\textwidth]{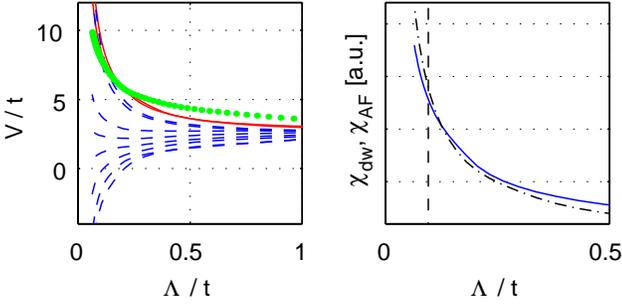}
\caption{Flow of certain coupling functions and susceptibilities in the saddle
point regime of the hole-doped case with $t'=-0.3t$ and
$\mu=-t$. The dashed lines in the left plot show the flow of Cooper couplings, 
the solid lines umklapp processes between the saddle points. The dotted
line denotes the largest coupling in the BZ diagonal. In the 
right plot the solid line denotes the $d$-wave pairing 
susceptibility $\chi_{d\mathrm{-wave}}$, the dashed-dotted line the AF susceptibility 
$\chi_s (\pi,\pi)$.}
\label{hdfvcomp}
\end{figure}
\end{center}

 The hole-doped case with $t'=-0.3t$ was extensively discussed in
Ref. \cite{honerkamp}. In order to simplify
the  comparison with results for the electron-doped case we repeat the main
observations briefly. 
We found  that in the density region where the FS crosses the US close to the saddle 
points the dominant processes which grow fastest upon integrating out higher
energy modes are given by mutually reinforcing  umklapp and Cooper processes
between the saddle point regions. This leads to a flow to strong coupling at
relatively high critical scales. The scattering processes involving
quasiparticles in the BZ diagonals grow as well but they are merely driven by the
couplings between the saddle points and diverge less rapidly. 
These results are shown in Fig. \ref{hdvcomp} and Fig. \ref{hdfvcomp}.  Starting the
flow with purely repulsive interaction, the Cooper channel will initially always
decrease the values of the Cooper scatterings. However if the FS is close to
the saddle points the particle-hole channel with the rapidly
growing umklapp processes also affects the large angle pair scattering
processes which involve momentum transfer $(\pi,\pi)$. In the saddle point
regime of the hole-doped case this effect is stronger than the suppression through
the Cooper channel and the  large angle pair scattering
processes grow upon reducing the energy scale which in turn enhances the flow
of the small angle pair scattering to negative values. Similarly it can be
also seen that the growth of the $d$-wave pair scattering enhances the flow of 
the umklapp $(\pi,\pi)$ processes. In the saddle point regime with the FS
close to the $(0,\pi)$ and $(\pi,0)$-points this mutual reinforcement between
$d$-wave and
AF processes is very effective as it only involves low-energy quasiparticles.
Therefore $d$-wave and AF processes diverge together at a single high energy scale. 

 \subsection{Flow in the electron-doped case}
\begin{center} 
\begin{figure} 
\includegraphics[width=.48\textwidth]{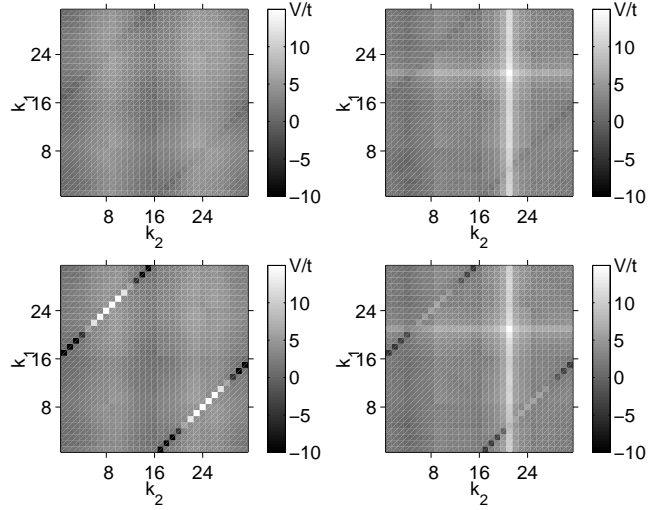}
\caption{Snapshot of the RG flow  when the couplings have reached values
larger than the bandwidth for the electron-doped case, $t'=-0.3t$, $T=0$. For
the upper two plots 
$\mu=-0.1t$  ($\langle n \rangle \approx 
1.18$), and the FS intersects the US. The left plots show the dependence of the scattering vertex
$V_\Lambda (k_1,k_2,k_3=1)$ with the first outgoing wavevector fixed at point 1 
on the incoming wavevectors $\vec{k}_F(k_1)$ and on $\vec{k}_F(k_2)$ where $k_i$
labels the positions of the points around the FS (see Fig. \ref{fshded}). In the 
upper right plot, $k_3=4$, i.e. the first outgoing wavevector is near the BZ
diagonal. The sharp vertical repulsive feature at $k_2=21$ is due to nesting with the
wavevector $\vec{k}_F(k_2) -\vec{k}_F(k_3) \approx (\pi,\pi)$. 
For the lower plots, $\mu=0.04t$ ($\langle n \rangle \approx 
1.22$) and the FS lies outside the US. 
The left plot again corresponds to $k_3=1$ and the right plot is for $k_3=4$. The diagonal features belong to Cooper
processes with $\vec{k}_F(k_1) +\vec{k}_F(k_2) =0$. The sign structure of the
Cooper pair scattering reveals that the dominant pairing symmetry is
$d_{x^2-y^2}$, e.g. $V_\Lambda (1,17,1)$ is attractive while $V_\Lambda
(1,17,9)$ is repulsive.}
 \label{edvcomp}
\end{figure}
\end{center}

\begin{center} 
\begin{figure} 
\includegraphics[width=.48\textwidth]{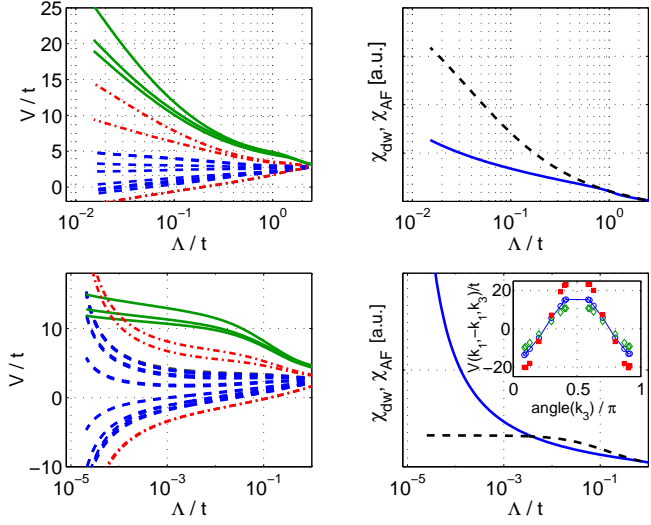}
\caption{Flow of coupling functions (left plots) and 
susceptibilities (right plots) in the
crossover regime of the electron-doped case, $t'=-0.3t$ and $T=0$. 
The upper plots correspond to $\mu=-0.1t$ where the FS intersects the US and 
the AF processes processes
diverge at non-zero energy scale. The lower plots are for $\mu =
0.04t$, which is further away from half-filling. 
The dashed lines in the left plots show the flow of Cooper couplings at the FS, 
the solid lines umklapp processes in the BZ diagonal $V_\Lambda (4,4,21)$,
$V_\Lambda (3,4,21)$, and $V_\Lambda (5,4,21)$. The dashed-dotted
lines denote Cooper processes between the saddle point regions with band
energies below the FS (discussed in the text).
 In the 
right plots the solid line denotes the $d$-wave pairing 
susceptibility and the dashed-dotted line the AF susceptibility 
$\chi_s (\pi,\pi)$. Note the logarithmic scale axis. The inset in the lower
right plot shows the Cooper pair scattering
$V(\vec{k}_1,-\vec{k}_1,\vec{k}_3)$ close to the instability as 
function of the angle of $\vec{k_3}$ around the FS. The filled squares (open diamonds)
correspond to scattering between $\vec{k}$-space points below (above) the FS at band
energy $\pm 0.8t$ while the circles connected by straight solid lines show the values
at for pairs on the FS.  $\vec{k}_1$ is at angle $\approx -\pi$ below (above)
or on the FS (see Fig. \ref{geom}).
}
\label{edfvcomp}
\end{figure}
\end{center}

For the electron-doped side we again take $t'=-0.3t$. Here, unlike the hole-doped
case, smaller absolute values of $t'$ yield a similar picture, only the critical scales become larger due to 
the improved nesting.  Half-filling corresponds to $\mu =-0.65t$. Upon
increasing  $\mu$ and $\langle n \rangle$, the
FS crosses the US close to the BZ diagonals until $\mu=0t$ and then loses contact to the US 
if we  further increase the particle number. At these densities the flow to
strong coupling changes drastically, therefore we concentrate on this filling range.
In Fig. \ref{edfvcomp} we show the flow to strong coupling for two different
chemical potentials: in the upper plots $\mu=-0.1t$ and the FS intersects the 
umklapp surface close to the BZ diagonals. Here the dominant 
processes which become large at comparably high energy scales 
are umklapp processes between the BZ diagonals involving the
momentum transfer $(\pi,\pi)$ (see solid lines in Fig. \ref{edfvcomp} and the sharp features in
Fig. \ref{edvcomp}). 

The Cooper processes between particles at the FS have dominant
$d_{x^2-y^2}$-symmetry but remain of the order of the bandwidth even when we
integrate the flow far out of the perturbative range. Thus the coupling to the strongly growing AF processes is only weak, as expected
from the location of the FS. Consequently the characteristic energy scales
where AF and $d$-wave channels start to grow are very different, in strong
contrast to the saddle point regime of the hole-doped case. Nonetheless we can still find signs of a coupling
between $d$-wave and AF channel: if we consider Cooper processes for particles
at the saddle points below the FS we find a much stronger growth towards
lower scales (see dashed-dotted lines in  Fig. \ref{edfvcomp}). A large angle Cooper pair scattering between two inequivalent saddle points
involves momentum transfer $\approx (\pi,\pi)$ and is therefore enhanced by the AF
processes. But unlike the saddle point regime in the hole-doped case, 
the $k$-space regions where the coupling occurs are away from the FS and
therefore there is only a weak influence of this mechanism on the true
low-energy excitations. One should also note that in more realistic
calculations including self-energy effects the coupling through off-FS 
processes will be further reduced by the short lifetimes of the
quasiparticles in these regions.

Increasing the band filling such that the FS loses its intersection with the
US, the umklapp and other $(\pi,\pi)$-processes in the diagonals get cut off and saturate at low
scales (see solid lines in the lower left plot of  Fig. \ref{edfvcomp}). 
Thus the $d$-wave Cooper processes can finally diverge at very low scales (at
least two orders of magnitude lower than in the previous case). The Cooper
pair scattering is shown in in the inset of Fig. \ref{edfvcomp}.
Again the off-FS $d$-wave Cooper processes closer to the saddle points benefit more from
the AF scattering processes and grow more strongly than the Cooper processes
for pairs at the FS, but this difference becomes less with increasing particle
density away from half-filling. Apart from this side-effect  we emphasize that in this
low-scale $d$-wave regime only Cooper processes diverge. Umklapp and and
nesting processes are cut off and do not develop singularities.

\section{Flow of the susceptibilities}
Next we compare the flow of various susceptibilities. First we analyze the real 
parts of the static
$d$-wave pairing susceptibility $\chi_{d\mathrm{-wave}}$ and the AF susceptibility
$\chi_s(\pi,\pi)$. Moreover, as a kind of cross-check, we analyze the coupling 
to uniform static charge and spin fields for electrons at different points on the Fermi surface .

\begin{center} 
\begin{figure} 
\includegraphics[width=.48\textwidth]{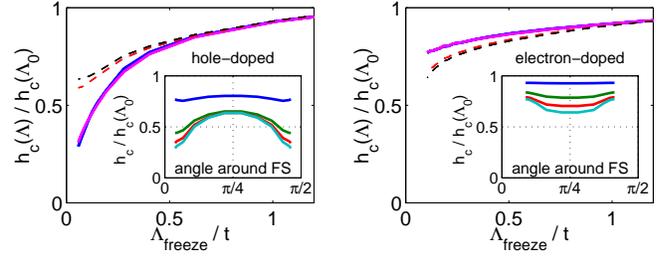}
\caption{Flow of the coupling of quasiparticles at different points on the FS to uniform external 
charge fields, normalized to the values with the bare interactions $U=3t$. The left plot corresponds
to the saddle-point regime of the hole-doped case $\mu=-t$, $T=0.04t$, while
the right plot shows data for the electron-doped case $\mu=-0.1t$,
$T=0.01t$. The solid lines are for $\vec{k}$-points on the FS in the vicinity
of the saddle points, the dashed lines for $\vec{k}$-points on the FS in the BZ
diagonal. $\Lambda_{\mathrm{freeze}}$ is the energy scale down to which the flow of the coupling function
is taken into account. Below $\Lambda_{\mathrm{freeze}}$ the coupling
functions are kept constant. The insets show the angular variation along the FS for
different values of $\Lambda_{\mathrm{freeze}}$. } 
\label{ucoupl}
\end{figure}
\end{center}

In the saddle point regime of the hole-doped case the overall
behavior of these probes together with general features of the RG flow of the
couplings indicated the tendency towards formation of an insulating spin liquid: both $\chi_{d\mathrm{-wave}}$ and
$\chi_s(\pi,\pi)$ diverge together in a similar way (see
Fig. \ref{hdfvcomp}) and in view of the mutual reinforcement between both
channels it is very plausible that the strong coupling state will
incorporate both types of fluctuations.
On the other hand the coupling of the saddle point
regions to uniform charge and 
spin fields becomes increasingly suppressed in the RG flow (the charge
couplings are shown in the left plot in
Fig. \ref{ucoupl}), suggesting the opening of charge and spin gaps around the
saddle points. In contrast to that the charge coupling for quasiparticles in
the BZ diagonals does not renormalize to zero. This anisotropic flow is
consistent with a
truncation of the FS at the saddle points while the BZ diagonals remain metallic.
The flow of the scattering rates described in Sec. \ref{scatrat} corroborates 
this scenario.

In the electron-doped case the situation is rather different. Since $d$-wave and AF 
channel do not couple strongly the corresponding susceptibilities have very
different flows. While $\chi_s(\pi,\pi)$ already grows considerably at higher
scales and eventually diverges if the system is not far away from half
filling, $\chi_{d\mathrm{-wave}}$ becomes large only at low scales. We therefore expect
that the system will have only two different phases if we allow for a
sufficient amount of 3D coupling: one AF phase with higher critical scale and 
temperature, and, if we increase the band filling such that the instability in
the latter channel becomes cut off by the FS curvature, a rapid drop in the energy
scale with a low-$T$ Kohn-Luttinger-type $d$-wave phase.  This -- compared to the 
saddle point regime -- more
conventional picture is also supported by the behavior of the uniform
susceptibilities and the $k$-space dependence of the coupling to the
corresponding external fields:
the coupling to uniform charge fields (see right plot in Fig. \ref{ucoupl}) is not strongly suppressed, i.e. there are no 
pronounced indications for incompressible or truncated parts of the FS in our
weak-coupling analysis. The reduction of the charge couplings becomes even
weaker when the filling is increased.
In contrast to the hole-doped case the strongest suppression now occurs for
wavevectors in the BZ diagonals, in agreement with the fact that Umklapp
processes flow most strongly there 
(see inset in the right plot of Fig. \ref{ucoupl}). Again this type of
anisotropy compares well the angular-dependent quasiparticle lifetimes
obtained with the RG calculations described below and FLEX (fluctuation
exchange approximation) \cite{kontani}.  

\section{Tentative phase diagram for the electron-doped case}
The tentative phase diagram for the hole-doped Hubbard model extracted from the one-loop flow 
to strong coupling was discussed in \cite{honerkamp}. Here
present a schematic phase diagram for the electron-doped side. 

\begin{figure}
\begin{center} 
\includegraphics[width=.48\textwidth]{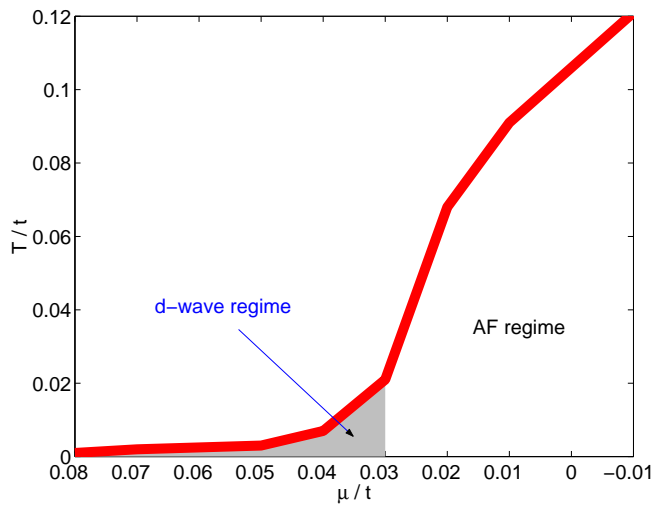}
\end{center} 
\caption{Dependence of the flow to strong coupling on $\mu$ and $T$ in the
electron-doped regime for $t'=-0.3t$ and $U=3t$. The thick line denotes the critical temperature above
which  we can integrate down to $\Lambda=0$ with all couplings
remaining smaller than 15$t$. In the $d$-wave dominated regime $\chi_{d\mathrm{-wave}}$ grows 
stronger, while in the AF regime $\chi_s(\pi,\pi)$ dominates (the precise
criteria are given in the text). For the displayed values of $\mu$ the
electron-doping varies between $0.2$ ($\mu=-0.1t$) and $0.23$ ($\mu=0.08t$).}
\label{pded}
\end{figure}

Generally with this type of calculations it is difficult to determine exact
phase boundaries, e.g. the parameters where the 
ground state changes from AF to $d$-wave superconducting long range order. Nor 
can we actually prove the existence of the phases suggested by one-loop
flow. Selfenergy and higher order effects will become large once the coupling functions exceed the order of the band width and from the flow below that scale it is
impossible to tell whether the different classes of couplings really diverge
or not. 
Similarly fluctuations around possible ordered states are not taken into account properly.
Nevertheless since in the electron-doped case we observe two different
tendencies which are weakly coupled and dominate on opposite sides of the
considered density range, we can well discriminate between distinct regimes with
different dominant fluctuations. If long range order is possible (at $T=0$ or
at finite $T$ if we add some degree of three-dimensional coupling), its type
will most probably correspond to the dominant fluctuations visible in the
one-loop RG.  
Bearing in mind these remarks, we present a schematic phase diagram
for the case $t'=-0.3t$ and initial $U=3t$ in Fig. \ref{pded}. 
For chemical potential $\mu < 0.03t$
(which corresponds to $\langle n \rangle \approx 1.21 $ per site) we 
find an AF dominated instability with comparably high critical temperature
$T_c$ (above $T_c$ we can integrate down to $\Lambda=0$ with all
couplings staying smaller than 15$t$.). For larger chemical potential the instability is 
$d$-wave dominated. The criterion we used for the distinction was the
derivative of  $\chi_{d\mathrm{-wave}}$ and $\chi_s(\pi,\pi)$ with respect to
$\Lambda$ when the maximal
coupling grows larger than 12$t$. We repeat that this distinction does not
allow any quantitative predictions for the true phase diagram. In particular
the critical scales and temperatures for the $d$-wave phase strongly depend on the
criterion used to separate the regimes\footnote{If we chose to compare the
susceptibilities at higher values of the couplings, i.e. $20t$ instead of $12t$, the $d$-wave phase would
extend further to smaller electron-dopings (to approx. $\mu=0.01t$). Very close to the instability
however, when the coupling function becomes huge, our one-loop RG scheme cannot be justified.}.

In the electron-doped case considered here there are two disparate energy
scales. The first one, which can be high close to half filling, is related to the AF-$(\pi,\pi)$ processes between the
BZ diagonals. The second energy scale is given by the critical scale of the
$d$-wave Cooper channel and remains much lower close to half-filling. Only if
the AF energy scale becomes too low at increased doping, when the FS curvature
prevents a real flow to strong coupling of the
AF processes, an instability in the $d$-wave Cooper channel can occur.  In the 
present case, the crossover energy scale is mainly determined by the particle density
which regulates the overlap of the  FS with the US,
i.e. the low energy phase space available for elastic scattering with momentum
transfer $(\pi,\pi)$.  A qualitatively similar picture was found by Manske et
al. \cite{manske} who solved the generalized Eliashberg equations with a spin-fluctuation
induced pairing interaction obtained within a FLEX scheme. 
 
As mentioned
above, on the electron-doped side the flow to strong coupling is qualitatively 
similar for all $t'<0$, only the critical scales and relative widths of the
regimes change.  Thus for band fillings larger than one particle per 
site our observations for $t'=-0.3t$ are in full qualitative agreement with
the results of Zanchi and Schulz \cite{zanchi} for $t'=0$ and
Halboth and Metzner \cite{halboth,hmprl} for smaller absolute values of the next-nearest 
neighbor hopping $t'$. In particular in the analysis of 
Ref. \cite{hmprl} the $d$-wave channel
becomes dominant roughly when the FS loses its intersection with the 
US upon increasing the electron density.

\section{Scattering rates}
\label{scatrat}
\subsection{Hole-doped case}
Here we describe our results for the quasiparticle scattering rates in the
hole-doped case. There we are mainly interested in electron densities
corresponding to the saddle point regime \cite{honerkamp}  briefly described above,
because there the flow to strong coupling becomes qualitatively different from the
$d$-wave superconducting instability at lower densities. Again, since we
cannot prove the FS truncation suggested by the flow in the saddle point
regime, our main goal is to analyze whether the behavior of the scattering
rates is consistent with this scenario.

\begin{figure}
\begin{center} 
\includegraphics[width=.5\textwidth]{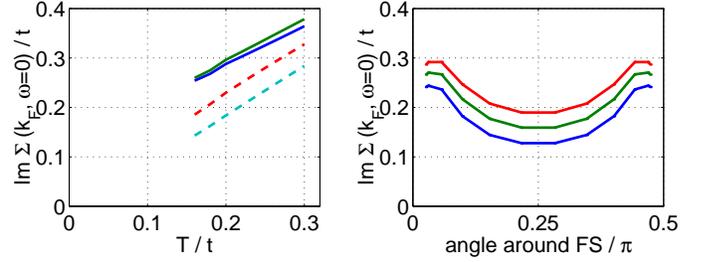}
\end{center} 
\caption{Left: Temperature dependence of imaginary part of the selfenergy at four
different positions on the FS for the hole-doped case with $\mu=-1.1t$ and $t'=-0.3t$. 
The dashed lines belong to points in the BZ
diagonal, the solid lines to points closer to the saddle points. Right: Imaginary part of the selfenergy versus angle around the FS in the saddle point regime at
$\mu=-1.1t$ for different temperatures above the instability. The lower line
(circles) corresponds to $T=0.15t$, the middle line (diamonds) to $T=0.18t$ and 
the upper line is at $T=0.21t$.  For $T=0.15t$ the largest coupling at $\Lambda=0$ is $\approx 11.5t$,
for $T=0.21t$ they reach $\approx 8.9t$. The saddle points are at angles 0 and
$\pi/2$.}
\label{ims110}
\end{figure}

Typical results for the scattering rates in the saddle point regime ($\mu=-1.1t$) above the critical
temperature obtained with the method described in 
Subsec. \ref{scatratform} are shown in Fig. \ref{ims110}. The temperature was
chosen such that the couplings $V_\Lambda$ do not become too large when we
integrate the flow to $\Lambda=0$, for the lowest $T$ shown the
maximum coupling reaches $\approx 11.5t$. This is already larger than the
bandwidth, but this high value is only reached at low scales and most
couplings remain smaller. Therefore the results should be qualitatively
correct. 
An anisotropy in the scattering rate is clearly observable but not
too pronounced, the maximum ratio between Im $\Sigma$ at the saddle points and
in the diagonals at $\pi/4$ is $\approx 2$. 
The temperature dependence of Im $\Sigma(\vec{k}_F, \omega =0)$ is shown in
Fig. \ref{ims110} for $\mu=-1.1t$ which is above the saddle point regime
and close to the van Hove doping at $\mu=-1.2t$. All curves for the different
positions on the Fermi surface show an almost linear increase with
$T$.  Both, the angular variation and the temperature dependence, are qualitatively consistent with the ARPES
results by Valla et
al. \cite{valla}\footnote{Valla et al. \cite{valla} find a leveling off of the scattering rate very close to the saddle points at lowest $T$ in the normal state, which is not
present in our results. If this is an intrinsic effect of the CuO$_2$-planes, it may be beyond our weak coupling calculation.}. In comparison with the model
assumptions in \cite{ioffe} we note that our calculations at temperatures
above the flow to strong coupling yield a smaller
anisotropy in the scattering rates and no $T^2$ behavior for the
quasiparticles in the BZ diagonal. Our data resemble more FLEX results
obtained by Kontani et
al. \cite{kontani} Altmann et al. \cite{altmann}. The first group was able to
describe resistivity and Hall coefficients in qualitative agreement with the experiments.
For high temperatures the qualitative similarity between FLEX calculations and our RG
analysis is not accidental because for $T>T_c$ when the flow does not diverge, the main scattering occurs in the 
$(\pi ,\pi)$ spin fluctuation channel which the typical FLEX schemes focus on.

From a theoretical point of view the linear-$T$ dependence 
is not unexpected because already the bare two-loop diagram with
unrenormalized couplings yields this behavior if the band filling is
sufficiently close to the van Hove filling \cite{lee,kastrinakis}. 
Hence it should be considered as
an effect of the van Hove singularities and does not immediately imply a
breakdown of the quasiparticle concept 
. 

\begin{figure}
\begin{center} 
\includegraphics[width=.5\textwidth]{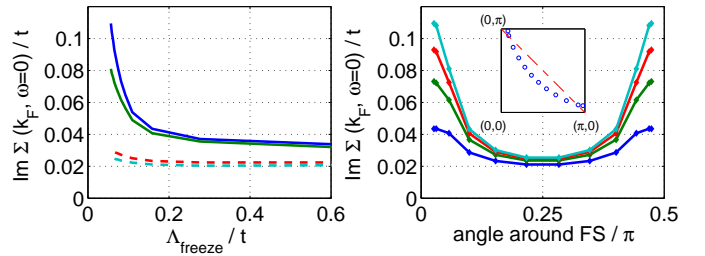}
\end{center}
\caption{Left: Change of the scattering rate through the flow of the interactions
$V_\Lambda$ for different positions on the Fermi
surface in the hole-doped case, calculated on a $48 \times 3 $ system. $\Lambda_{\mathrm{freeze}}$ is the scale below which the interactions
are kept constant. Right: Angular variation of Im $\Sigma (\vec{k}_F,
\omega=0)$ with $\vec{k}_F$ varying on the FS from one to the next saddle
point. The different  lines shows $\Sigma (\vec{k}_F,
\omega=0)$ with the couplings stopped at $V_{\Lambda, \mathrm{max}} = 6t$
(bottom line), $9t$, $12t$ and $15t$ (top line). The inset shows the 12 points on 
a quarter of the FS. For both plots $\mu=-1.1t$ and $T=0.04t$ corresponding to 
the saddle point regime.}
\label{imsla}
\end{figure}

For lower temperatures the coupling function flows to strong coupling at a non-zero
critical scale $\Lambda_c$. Therefore we cannot integrate the flow down to
zero scale and do not obtain a good approximation for 
Im $\Sigma$ with the above method.  At $\Lambda_c$ the selfenergy diverges
together with the couplings and we expect that the quasiparticles will be at
least partially destroyed. Although we cannot really access this strong
coupling region, we can gain some insight on how this
quasiparticle destruction takes place by considering the change in the two-loop
selfenergy through the flow of the coupling function.
 The results for $\mu=-t$ and $T=0.04t$ are shown in
Fig. \ref{imsla}. If we freeze the flow of the four point vertices $V_\Lambda$ already at high
scales, we basically obtain the bare two-loop selfenergy when we integrate
the RG equation for the selfenergy down to $\Lambda =0$. There, the anisotropy
between saddle point regions and BZ diagonals 
is small.  If we now subsequently include the flow of $V_\Lambda$ by reducing $\Lambda_{\mathrm{freeze}}$, the
scattering rate for quasiparticles around the saddle points grows
strongly. For the quasiparticles in the BZ diagonals the scattering rate is not much 
affected by the flow of the couplings.  Thus the flow of the scattering rates  for these
values of the band filling is consistent with a breaking up of 
the FS into two distinct regions, as suggested in \cite{honerkamp} based on an 
analysis of the flow of the coupling function and susceptibilities: 
around the saddle points, the quasiparticles
are subject to divergent scattering processes, there we also found the strong
suppression of the charge compressibility. In the BZ diagonals which appeared to
remain compressible, the scattering rate stays in the weak coupling range, consistent
with the FS remaining untruncated in these regions.

When the band filling is increased further towards half-filling the energy and
temperature scale for the flow to strong coupling rises and the anisotropy of
the scattering rates at temperatures above this flow to strong coupling
becomes weaker. The reason is that the scattering processes between
the saddle point regions lose their
dominant role, and the strongest scattering processes occur between FS parts
connected by $(\pi,\pi)$, now closer to the BZ diagonals. 

\subsection{Electron-doped case}
\begin{figure}[h]
\begin{center} 
\includegraphics[width=.5\textwidth]{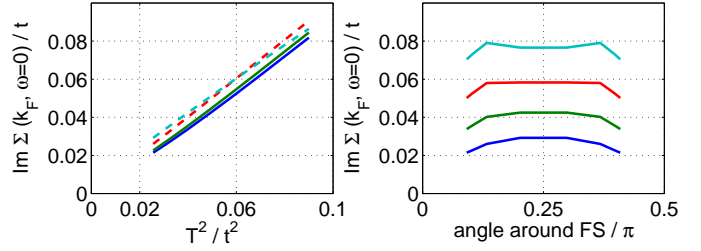}
\end{center} 
\caption{Left: Temperature dependence of imaginary part of the selfenergy at four
different positions on the FS for the electron-doped case. The dashed lines belong to points in the BZ
diagonal, the solid lines to points closer to the saddle points. Right:
Angular variation of Im $\Sigma (\vec{k}_F,
\omega=0)$ with $\vec{k}_F$ varying on the FS from one to the next saddle
point. The different  lines show $\Sigma (\vec{k}_F,
\omega=0)$ at different temperatures from $T=0.16t$
(bottom line) to $T=0.28t$ (top line).
The results in both plots were obtained using a $32\times 3$ system with $t'=-0.3t$ and $\mu=0$.}
\label{ims0Tdep}
\end{figure}

\begin{figure}
\begin{center} 
\includegraphics[width=.5\textwidth]{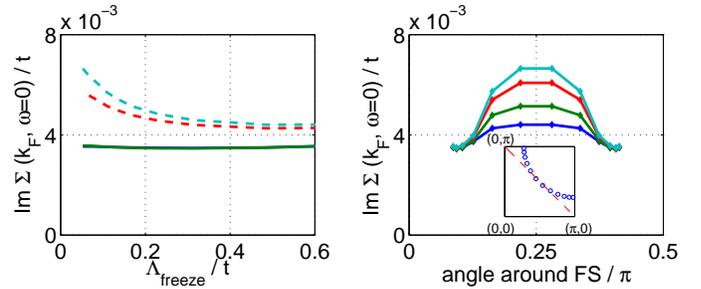}
\end{center}
\caption{Left: Change of the scattering rate through the flow of the interactions
$V_\Lambda$ for different positions on the Fermi
surface in the electron-doped case. $\Lambda_{\mathrm{freeze}}$ is the scale below which the interactions
are kept constant. Right: Angular variations of Im $\Sigma (\vec{k}_F,
\omega=0)$ with $\vec{k}_F$ varying on the FS from one to the next saddle
point. The different  lines shows $\Sigma (\vec{k}_F,
\omega=0)$ with the couplings stopped at $V_{\Lambda, \mathrm{max}} = 6t$
(bottom line), $9t$, $12t$ and $15t$ (top line). The inset shows the 12 points on 
a quarter of the FS. For both plots $\mu=-0.1t$ and $T=0.04t$.}
\label{imsla10}
\end{figure}

In Fig. \ref{ims0Tdep} we show the results for the electron-doped case 
obtained at higher temperatures
where the flow does not exceed the order of the bandwidth. There are 
striking differences compared to the results obtained in the hole-doped saddle point
regime. First the overall magnitude of Im $\Sigma (\vec{k}_F,\omega=0)$ at a
given temperature is much smaller, almost by an order of magnitude. We
attribute this to the reduced density of states around the
FS, the van Hove singularities are far away from the FS. Second the
temperature dependence of the scattering rates is far from being linear and appears to be roughly consistent with 
a $T^2$-behavior at low temperatures above the flow to strong
coupling\footnote{In an isotropic two-dimensional Fermi liquid, we expect Im $\Sigma
(\omega=0,\vec{k}_F) \sim (T/T_F)^2 \log (T_F/T)$ \cite{chaplik,hodges}. Since 
our calculations are rather qualitative we do not attempt a detailed
comparison of the data with
this $T$-dependence.}. The data for $\vec{k}$-points in the BZ diagonal do not extrapolate
to 0 for $T\to 0$ due to the increase of the coupling function at lower $T$.
Further, the anisotropy of the scattering rate is different and much weaker 
than in the hole-doped case. The FS parts which feel the strongest scattering
at lower $T$ are now in the BZ
diagonal (see right plot in Fig. \ref{ims0Tdep} for angular dependence). This
is consistent with the flow of the interactions which grow fastest in
these regions. Our results also agree qualitatively with FLEX calculations \cite{kontani,altmann},
which show
that the  in the electron-doped case the ''hot spots'' with the shortest
quasiparticle-lifetime move towards the BZ diagonal. Nonetheless due to the
weakness of the anisotropy the data do not suggest a truncation of the FS like 
in the hole-doped case. This can be also seen at lower temperatures (see
Fig. \ref{imsla10}), where flow of the scattering rates remains comparably
weak  when the coupling function flows to 
strong coupling.
\section{Discussion}
We have presented $N$-patch RG results for the two-dimen\-sional $t$-$t'$ Hubbard model
with particle density $n>1$ per site and $t'=-0.3t$. We have shown that the RG 
flow to strong coupling is qualitatively different from the $n<1$ case because of
the different location of the FS in both cases. The saddle-point regime 
\cite{honerkamp} which we
associated with the pseudogap state in the (hole-)underdoped cuprates is absent 
in our RG study of the electron-doped side. Instead the picture closely
resembles the observations made by Zanchi and Schulz \cite{zanchi} and
Halboth and Metzner \cite{halboth,hmprl} for zero or smaller absolute value of the
next-nearest neighbor hopping $t'$. 
Upon increasing the electron density we find a rapid decrease of the 
critical scale for the instability where the flow of the otherwise strongly
growing AF susceptibility is cut off. This leaves a small window for a 
Kohn-Luttinger-type $d_{x^2-y^2}$-wave instability at lower scales and
temperatures.
The tendency towards incompressibility and truncation of parts of the FS is weaker than in the hole-doped case and strongest in the BZ diagonal. 
At temperatures above the strongly coupled phase we find
slightly anisotropic scattering rates with strongest scattering of
quasiparticles, i.e. ''hot spots'' in the BZ
diagonal. The temperature dependence of the scattering rates is roughly quadratic.

Indeed the experimental phase diagram of the electron-doped cuprates shows a
relatively wide AF phase with a boundary to superconducting phase with
lower 
critical temperature. There is increasing experimental evidence \cite{tsuei,prozorov,mourachkine,armitage} that the pairing symmetry
of this phase is $d_{x^2-y^2}$. Moreover the in-plane resistivity in the
normal state shows a quadratic temperature dependence \cite{tsueigupta,peng}.

Note that in our RG study the difference between 
the energy scales characteristic to these phases is considerably larger than the experimental 
values. However one should keep in mind that our one-loop analysis can only provide
 a qualitative picture for the understanding of the real materials. In particular we
cannot make any predictions about the stability of the phases suggested 
by the RG flow and the precise location of the transition from AF to $d$-wave
regime. In fact the Fermi surface in 
Nd$_{1.85}$Ce$_{0.15}$CuO$_4$ with a superconducting $T_c$ of 24$K$ seen by ARPES \cite{armitage} still intersects the Umklapp surface. Thus it might 
be that our criterion for the transition from AF to $d$-wave
regime overestimates the width of the AF regime.

The saddle point 
regime in the hole-doped case with the diverging umklapp processes was interpreted as a
precursor of the Mott state where the FS becomes truncated at the saddle
points \cite{honerkamp}. The anisotropic scattering
rates described above, 
which grow strongest at the saddle points as the flow goes to strong
coupling, are consistent with this scenario. 
Due to the location of the FS the diverging umklapp processes also
generate strong $d$-wave pairing correlations. In this situation the
$d$-wave-paired condensate which may form at sufficiently low temperature will
still be subject to the umklapp scattering at the saddle points where the FS
becomes truncated. This may result in a reduced superfluid weight arising from the open
parts of the FS in the BZ diagonals only. 

In the electron-doped case the umklapp and other AF processes predominantly occur between the BZ
diagonals and do not strongly drive the $d$-wave correlations, they can only 
lead to a rapid growth  of AF correlations. Thus,
instead of being strongly coupled and mutually reinforcing each other in the flow, AF and $d$-wave 
channel have disparate characteristic energy scales and influence each other
only weakly.
As soon as the umklapp  processes in the BZ diagonals flow to strong coupling,
the instability is entirely dominated by AF
correlations because the energy scale of the $d$-wave instability is much
lower. 
This implies that coming from the
electron-overdoped side the precursors
of the Mott state only occur in the AF phase and should have less influence on 
the $d$-wave state. In particular we speculate that in the superconducting
state all
electrons contribute to the superfluid weight in contrast with the hole-(under)doped
case. 
\\[4mm]

It is a pleasure to thank T.M. Rice, M. Salmhofer, U. Ledermann, K. Lehur,
M. Zhitomirsky,  R. Hlubina and 
D. Vollhardt for stimulating discussions. The Swiss National Science Foundation is
acknowledged for financial support. The numerical calculations were performed
on the {\em Asgard} Beowulf cluster of ETH Z\"urich.

\end{document}